\documentclass[preprint,aps, preprintnumbers, nofootinbib]{revtex4}
\usepackage[dvips]{graphics}

\newcommand{\lsim}[1]{
\setlength{\unitlength}{12pt}
\begin{picture}(1.4,1.)
\put(.7,-0.3){\makebox(0.0,1.)[t]{$<$}}
\put(.7,-0.3){\makebox(0.0,1.)[b]{$\sim$}}
\end{picture}#1}
\newcommand{\gsim}[2]{
\setlength{\unitlength}{12pt}
\begin{picture}(1.4,1.)
\put(.7,-0.3){\makebox(0.0,1.)[t]{$>$}}
\put(.7,-0.3){\makebox(0.0,1.)[b]{$\sim$}}
\end{picture}#2}

\begin{document}

\title{On Internal Consistency of Holographic Dark Energy Models}

\author{R. Horvat}
\email{horvat@lei3.irb.hr}
\address{Rudjer Bo\v{s}kovi\'{c} Institute, P.O.B. 180, 10002 Zagreb,
Croatia}

\begin{abstract}

Holographic dark energy (HDE) models, underlain by an effective quantum field
theory (QFT) with a manifest UV/IR connection, have become a convincing
candidate for the dark energy in the universe. On the other hand, the
maximum number of quantum states a conventional QFT in the box of size $L$ is 
capable to  describe, refer to those boxes which are 
on the brink of experiencing a 
sudden collapse to a black hole. Another restriction on the underlying QFT
is that the UV cutoff, which cannot be chosen independently of the IR cutoff
and therefore becomes a function of time in a cosmological setting, should
stay the largest energy scale even in  the standard cosmological epochs
preceding a dark energy dominated one. We show that, irrespective of whether
one deals with the saturated form of HDE or takes a certain degree of
non-saturation in the past, the above restrictions cannot be met in a
radiation-dominated universe, an epoch in the history of the universe which
is expected to be perfectly describable within  conventional QFT.         
\end{abstract}

\newpage

\maketitle

The holographic principle \cite{1, 2} is undoubtedly the most amazing
ingredient of a modern view of space and time. The most successful
realization of the holographic principle, implemented in the Malcadena's 
discovery of AdS/CFT duality \cite{3}, strengthened further this 
speculative idea about quantum gravity. On the other hand, other closely
related concepts forming crucial parts within a new 
paradigm, like black hole complementarity \cite{4}, the UV/IR
connection \cite{5}, the space-time uncertainty relation \cite{6} and
the occurrence of the minimal length scale \cite{7},  
do become completely manifest in the framework of the holographic principle.

In order to encode (via the holographic information) a drastic depletion of 
quantum states within the effective 
field-theoretical description, preventing at the same time formation of 
black holes, the entropy for an effective quantum
field theory $\sim $ $L^3 {\Lambda }^3 $, where $L$ is the size of the
region
(providing an IR cutoff) and $\Lambda $ is the UV cutoff, should obey the
upper bound \cite{8}
\begin{equation}
L^3 {\Lambda }^3 \leq (S_{BH})^{3/4} \sim L^{3/2} M_{Pl}^{3/2}\;,
\end{equation}
and $S_{BH} \sim L^2 M_{Pl}^2 $ is the holographic
Bekenstein-Hawking entropy. In an
expanding universe $\Lambda $ should therefore be promoted to a varying
quantity (some decreasing function of $L$), in order (1) not to  be violated 
during the course of the
expansion, manifesting thereby explicitly the UV/IR correspondence. This gives a 
constraint on the maximum energy density in the effective 
theory, $\rho_{\Lambda } \leq  L^{-2}M_{Pl}^2 $. Obviously, 
$\rho_{\Lambda }$ is the energy density corresponding to a zero-point 
energy and the cutoff $\Lambda $ \footnote{Indeed, this may be seen by 
calculating the
effective cosmological constant (CC) generated by vacuum 
fluctuations (zero point energies)
\begin{eqnarray}
\rho_{\Lambda } \; \propto \; \int_{L^{-1}}^{\Lambda }k^2 dk
\sqrt{k^2 + m^2} \; &\sim& \; \Lambda^4 \; \; \; \; \; \; \; \;
\Lambda \gsim \,\,\,   m
\nonumber \\
&\sim& m \Lambda^3 \; \; \; \; \; \;
\Lambda \lsim \,\,\,   m \;, \end{eqnarray}
since clearly $\rho_{\Lambda }$ (as withal the entropy) is dominated by UV 
modes. Although the constraint $\rho_{\Lambda } \leq  L^{-2}M_{Pl}^2 $ stays
the same for both limiting cases, the UV/IR
correspondence is however different, giving $L \sim
\Lambda^{-2}$ and $L \sim
\Lambda^{-3/2}$, respectively, when the bound is 
saturated. Such a distinction becomes important in a
subsequent discussion.}.   	

The origin of (1) stems from the fact that in an effective QFT the entropy
scales extensively, $S \sim L^3 \Lambda^3 $, and therefore (for any
$\Lambda $) there is a sufficiently large volume for which $S$ would exceed
the absolute bound $S_{BH}$. Thus, considerations for the maximum possible
entropy suggest that ordinary QFT may not be valid for arbitrarily large
volumes, unless the UR and IR cutoffs satisfy a bound, $L \Lambda^3 \lsim
M_{Pl}^2 $. However, when saturated, this bound means that an effective QFT
should also be capable to describe systems containing black holes, since 
it necessarily includes many states with Schwarzschild radius much larger 
than the box size. The arguments of why an effective QFT appears unlikely to 
provide an adequate description of any system containing a black hole can be
found in \cite{8} and in references therein. So, ordinary QFT may not be
valid for much smaller volumes, but would apply provided (1) is satisfied. 

The above effective field-theoretical  
setup has triggered a novel variable CC approach \cite{9} generically
dubbed `holographic dark energy' (HDE) \cite{10}. The main reason of why the above
HDE model is so appealing in possible description
of dark energy is  when the holographic bound (1) is saturated, 
$\rho_{\Lambda }$ gives the right amount of dark energy in the
universe at present, provided $L$ today is of order of the Hubble parameter.
Moreover, since $\rho_{\Lambda }$ is now a running quantity, it also
has a  potential  to shed some light on the `cosmic coincidence problem'
\cite{11}. In addition, the original model \cite{8} is capable to satisfy
current observations \cite{12}, since by construction it has
$\omega_{\Lambda } = -1$. 

The most pressing problem when dealing with cosmologies based on (1) is 
certainly the
choice for the infrared cutoff $L^{-1}$. For models based on  full 
saturation in (1), the choice in the form of   
the inverse Hubble parameter is
largely unsatisfactory (for spatially flat
universes as suggested by observations) 
both for perfect fluids \cite{9} as well as
interacting fluids \cite{13, 14}; in the former case one cannot explain the
accelerating expansion of the present universe, while one fails to explain
that the acceleration sets in just recently and was preceded by a 
deceleration era at $z \gsim ;1$, in the latter case. This is easy to see by
plugging $\rho_{\Lambda } =  L^{-2}M_{Pl}^2 $ (setting a prefactor to unity
for simplicity) into the Friedman equation for flat space
\begin{equation}
(HL)^2 = \frac{8 \pi }{3} (1 + r)\;,
\end{equation}
where $r=\rho_m /\rho_{\Lambda}$ and $\rho_m $ is the matter 
energy density. Thus, a choice $L \sim H^{-1}$ would
require the ratio $r$ to be a constant. This is a general statement, holding
irrespective a fluid is
perfect or not, even irrespective the Newton constant is varying or not. The 
interpretation for
various cases is, however, different. For perfect fluids, $r=const.$ means
that the equation of state for the dark energy unavoidably  matches that of
pressureless matter, $w =0$ \cite{9}. Thus, we cannot explain the
accelerating
expansion of the present universe. For interacting fluids, one is usually
able to
generate accelerated expansion with $r=const.$ as now $a \sim t^{2/3}$
switches to $a \sim t^{2/3m}$, where the parameter $m$ depends on the
interaction term and can be easily made $m < 2/3$ \cite{13, 14} so as to
ensure acceleration. The constancy of $r$ for the flat space case 
precludes however any transition
between the cosmological epochs. Generically, a
suggestion of setting $L$ according to the future event horizon \cite{15}
leads to phenomenologically viable models. Even the choice $L=H^{-1}$ can be
saved in models where a certain degree of non-saturation in (1) is allowed
in the past \cite{14, 16}. 

In the present paper, we aim to check whether the effective
field-theoretical setup underlying (1), is capable of describing
various cosmological epochs  consistently. That is, if (1) , besides the
late-time acceleration in a dark-energy dominated epoch, provides also a
consistent description of some earlier epoch, say the radiation-dominated
universe. The consistency check will be based on the following two
requirements: ($\sl i $) a radiation-dominated epoch is considered as a system
at a temperature $T$, having thermal energy $L^3 T^4 $, provided $L^{-1} < T
< \Lambda(L)$, where the UV cutoff $\Lambda(L)$ is now a running quantity
which is to comply with (1), and ($\sl ii $) the range of validity of the effective
QFT is restricted only to those systems not containing black holes.  
In the following we argue that as soon as we move from the
epoch at which the dark  energy overwhelmingly dominates all other forms of
energy densities, any consistent description based on (1) is no longer 
viable \footnote{It was already noted in the background paper \cite{8} that
within an
effective QFT, requiring cutoffs which do obey (1), there is no
admissibility of
simultaneously addressing the CC problem and complying with computations
relevant for
current laboratory experiments. To address the CC problem with a naive
estimate, $\rho_{\Lambda } \sim \Lambda^4 $, one requires a UV cutoff of
order of $10^{-2.5}$ eV. Such a cutoff would induce a discrepancy in the
calculation of $(g-2)$ for the electron, between a framework  relying on
(1) and a conventional
one performed in an infinite box, which is unacceptably large, c.f. Eq.(5)
in \cite{8}. Here we focus on consistent description within
cosmology only.}. This conclusion remains irrespective of the choice 
for $L$ and a degree of saturation in (1). 

Let us begin with models saturating (1). As mentioned earlier, apart from a
small set of models presented in \cite{14, 16}, the bulk of the models 
have employed (1)
in its saturated form. The HDE density in the latter case is  
conveniently parametrized as
$\rho_{\Lambda } =(3c^2/8 \pi ) L^{-2} M_{Pl}^2$ \cite{10} , with a 
parameter $c^2 $ of order of unity. The corresponding
Schwarzschild radius for a box of volume $L^3 $ dominated by $\rho_{\Lambda
}$,  
\begin{equation}
R_{s} \sim M_{Pl}^{-2} (L^3 \rho_{\Lambda }) \sim L \;,
\end{equation}      
always sets the system at the brink of collapse to a black hole. It is
important to note that this contribution of $\rho_{\Lambda }$ in $R_{s}$ is
{\sl always} such throughout the history of the universe, notwithstanding of which
form of energy dominates a particular epoch. This means that every epoch
preceding a dark energy dominated phase, where $\rho_{\Lambda }$
necessarily represents a subdominant component in the total energy density of
the universe, would set $R_{s}$ at $R_{s} >> L$. That is, in epochs when
$\rho_{\Lambda }$ is subdominant, $R_s $ would rise up, $R_s >> L$,
since $R_s $ is determined by the ${\it total}$ energy density. That way, if
the same effective QFT is to describe other cosmological epochs besides the
current one, it necessarily includes many states with $R_s $ much larger
than the box size. In order 
not to contradict standard
cosmology, one should assume that in a radiation-dominated universe
$\rho_{rad} \sim T^4$, and thus with $\rho_{rad} >> \rho_{\Lambda }$ one
necessarily includes many states with $R_{s} >> L$, which are not expected
to be describable within conventional QFT. One can be hoped to remedy the
situation by applying the same recipe leading to (1), which brings HDE to a
domain describable within QFT. This amounts that the total energy of the
system of size $L$ should not exceed the mass of the the same-sized black
hole, i.e., $L^3 T^4 \lsim L M_{Pl}^2$. Apparently, radiation states would
be safe as now $R_{s} \lsim L$. The same constraint would however preclude
radiation from being the dominant 
component, as $\rho_{rad} \sim T^4 \lsim L^{-2} M_{Pl}^2 \sim \rho_{\Lambda }$.
So, we see that any saturated HDE model precludes either description of
radiation within ordinary QFT  or a transition between the cosmological
epochs \footnote{It is remarkable to note that even the
absolute Bekenstein-Hawking bound can be saturated in the radiation-dominated
epoch. Take for instance the popular Li's model \cite{10} and solve for 
$\rho_{rad} >> \rho_{\Lambda }$. One obtains, $\rho_{\Lambda } \simeq
\rho_{rad0} a^{-3}$, where the subscript  `0' denotes the present-day value.
This determines, in turn, the IR cutoff as $L \sim M_{Pl}
(\rho_{rad0})^{-1/2} a^{3/2}$. Equipped with these 
relationships, and $T \sim a^{-1}$, we find
that the absolute bound $L^3 T^3 \lsim L^2 M_{Pl}^2 $, is saturated at $T
\sim 10^9 $ GeV. This gives an interesting limit on the post-inflation
reheating temperature, i.e. the temperature at the beginning of the hot big
bang universe. Since Li's model employs a saturated version of (1),
when the absolute bound is saturated we have $R_s >> L$, and therefore it
is clear that this interesting bound cannot be fully trusted.}.        
 
Arguably much better prospects can be expected for models consistent with
(1) but, at the same time, allowing a certain degree of non-saturation in
the past epochs. In these so-called non-saturated HDE models \cite{14, 16},
the parameter $c^2$ is promoted to a function of cosmic time, $c^2 (t)$. The
function $c^2 (t)$ should satisfy $c^2 (t_0 ) \rightarrow 1$ (dark energy
dominance), while $c^2 (t) << 1$ during the radiation-dominated epoch. 
A promising setup, in which
each epoch represents a system  not containing black holes would be 
\begin{eqnarray}
T^4 \sim c_{rad}^2(t) L^{-2} M_{Pl}^2 \;, 
\\
\rho_{\Lambda } \sim c_{\Lambda}^2(t) L^{-2} M_{Pl}^2 \;,
\end{eqnarray} 
where the holographic bound is saturated asymptotically only by $c_{\Lambda}^2
$ (in order to ensure the current dark energy dominance). On the other hand,
$c_{rad}^2 >> c_{\Lambda}^2 $ (also $c_{rad}^2 \lsim 1 $) 
in the radiation-dominated era. The setup as given by Eqs. (5-6), with $c^{2}s
\lsim 1$, is only a formal account of a system free of black holes. Note that 
with the choice (6) the UV/IR
correspondence becomes more complicated, now depending on the particular 
choices for $c_{\Lambda}^2$. The bottom line, however, is that, by UV/IR
mixing, $\rho_{\Lambda }$ should always
(irrespective of the form of $c_{\Lambda }^{2}$) acquire the form
given by Eq. (2) (footnote 1). The era of
radiation-dominance therefore imposes a constraint, $T^4 > \rho_{\Lambda }$. On 
the other hand, in
a conventional QFT with some infrared limitation, a system at a temperature
$T$ has an energy $L^3 T^4 $ (and therefore energy density $T^4 $), provided
$L^{-1} < T < \Lambda $. If the mass scale is negligible with respect
to the UV cutoff, $\rho_{\Lambda } \sim
\Lambda^4 $ (see footnote 1), the above constraints are 
impossible to satisfy simultaneously  
throughout the
radiation-dominated era, showing thus internal inconsistency. With a more
realistic estimate, $\rho_{\Lambda } \sim m \Lambda^3 $ (see footnote 1), 
one obtains
\begin{equation}
\Lambda > T > m^{\frac{1}{4}} \Lambda^{\frac{3}{4}} \;.
\end{equation}
Thus, when $m > \Lambda $,
Eq. (7) would entail [via the case b) of Eq. (2)] $m < \Lambda $, showing 
internal inconsistency again. We have thus seen that though
with some degree of non-saturation of the holographic bound, both
systems (dark energy and radiation) can be made free of states lying within
their Schwarzschild radius, a consistent description of takeover of the
dominance by radiation within the same QFT is not possible.

Perhaps the situation is even  worse than stated above. If an effective QFT
is to encompass the standard models particles $(m \gsim \~ 100 $ GeV), the
present-day UV cutoff is much smaller than $10^{-2.5}$ eV. Indeed, from $m
\Lambda_{0}^3 \sim 10^{-11}$ eV$^4$, one obtains $\Lambda_{0} \sim 10^{-7}
$ eV. This means that even in the present epoch (dominated by dark energy fluid) a
consistent description of (CMBR) radiation is not viable since 
the present temperature of the universe $T_0 \sim
10^{-4}$ eV. 
  
In conclusion, we have shown that an effective QFT, with a proposed
relationship between UV and IR cutoffs as to eliminate the need for
fine-tuning in the `old' cosmological constant problem and explain 
furthermore dark
energy at present, cannot  describe consistently a radiation-dominated 
universe. Such a framework is 
particularly compelling in description of an expanding 
universe 
since without a corresponding UV/IR mixing, conventional QFT may not
be valid for arbitrarily large volumes. Albeit in a
radiation-dominated epoch the UV/IR correspondence can be made virtually
arbitrary, takeover by radiation cannot yet be obtained. Our result are
quite generic in that they do not depend on the pressing problem of the
choice of the IR cutoff. Because of 
the absence of a prominent energy scale (connected to microphysics),
disparate mass scales as well as  possibility that  underlying framework
may not be QFT (i.e. a black-hole fluid), we cannot {\sl a priori} draw the 
same conclusion for a matter-dominated epoch. Still, on similar grounds as
above it is seen that saturated HDE models would compromise a consistent description of
that epoch as well. Our overall conclusion is  
therefore that the
basic framework underlying all HDE models seems too {\sl ad-hoc} 
to have any real explanatory value, which 
still keeps us in need of firmer theoretical background.

{\bf Acknowledgment. } This work was supported by the Ministry of Science,
Education and Sport
of the Republic of Croatia under contract No. 098-0982887-2872.

\end{document}